\documentclass[12pt,preprint]{aastex}
\usepackage{emulateapj5,apjfonts}
\usepackage{graphics}
\usepackage{amssymb}
\usepackage{onecolfloat}
\lefthead{Luki\'c, Reed, Habib, Heitmann}
\righthead{The Structure of Halos: Implications for Group and Cluster Cosmology}

\begin{document}


\def\head{
  \vbox to 0pt{\vss
                    \hbox to 0pt{\hskip 440pt\rm LA-UR-07-7956\hss}
                   \vskip 25pt}

\title{The Structure of Halos: Implications for Group and Cluster Cosmology}
\author{Zarija~Luki\'c\altaffilmark{1,2},
	Darren~Reed\altaffilmark{3,4},
        Salman~Habib\altaffilmark{2}, and         
        Katrin~Heitmann\altaffilmark{4}}

\affil{$^1$ Dept.\ of Astronomy, University of Illinois, Urbana, IL 61801}
\affil{$^2$ T-8, Theoretical Division, Los
Alamos National Laboratory, Los Alamos, NM 87545}
\affil{$^3$ T-6, Theoretical Division, Los
Alamos National Laboratory, Los Alamos, NM 87545}
\affil{$^4$ ISR-1, ISR Division, Los
Alamos National Laboratory, Los Alamos, NM 87545}

\date{today}

\begin{abstract}
  The dark matter halo mass function is a key repository of
  cosmological information over a wide range of mass scales, from
  individual galaxies to galaxy clusters. N-body simulations have
  established that the friends-of-friends (FOF) mass function has a
  universal form to a surprising level of accuracy ($\lesssim
  10\%$). The high-mass tail of the mass function is exponentially
  sensitive to the amplitude of the initial density perturbations, the
  mean matter density parameter, $\Omega_{m}$, and to the dark energy
  controlled late-time evolution of the density field. Observed group
  and cluster masses, however, are usually stated in terms of a
  spherical overdensity (SO) mass which does not map simply to the
  FOF mass. Additionally, the widely used halo models of structure
  formation -- and halo occupancy distribution descriptions of
  galaxies within halos -- are often constructed exploiting the
  universal form of the FOF mass function. This again raises the
  question of whether FOF halos can be simply related to the notion of
  a spherical overdensity mass. By employing results from Monte Carlo
  realizations of ideal Navarro-Frenk-White (NFW) halos and N-body
  simulations, we study the relationship between the two definitions
  of halo mass. We find that the vast majority of halos ($80-85\%$) in
  the mass-range $10^{12.5}-10^{15.5}h^{-1}M_\odot$ indeed allow for an
  accurate mapping between the two definitions ($\sim 5\%$), but only
  if the halo concentrations are known. Nonisolated halos fall into
  two broad classes: those with complex substructure that are poor
  fits to NFW profiles and those ``bridged'' by the (isodensity-based)
  FOF algorithm. A closer investigation of the bridged halos reveals
  that the fraction of these halos and their satellite mass
  distribution is cosmology dependent. We provide a preliminary
  discussion of the theoretical and observational ramifications of
  these results.
\end{abstract}

\keywords{methods: N-body simulations ---
          cosmology: halo mass function}}

\twocolumn[\head]
\section{Introduction}

A large number of astronomical and cosmological observations now
provide compelling evidence for the existence of dark matter. Although
the ultimate nature of the dark matter remains unknown, its large
scale dynamics is completely consistent with that of a
self-gravitating collisionless fluid. In an expanding universe, the
gravitational instability is the driver of the growth of structure in
the dark matter, the final distribution arising from the nonlinear
amplification of primordial density fluctuations. The existence of
localized, highly overdense clumps of dark matter, termed halos, is an
essential feature of nonlinear gravitational collapse in cold dark
matter models.

Dark matter halos occupy a central place in the paradigm of structure
formation: Gas condensation, resultant star formation, and eventual
galaxy formation occur within halos. The distribution of halo masses
-- the halo mass function -- and its time evolution, are sensitive
probes of cosmology, particularly so at low redshifts, $z<2$, and high
masses. This last feature allows cluster observations to constrain the
dark energy content, $\Omega_{\Lambda}$, and the equation of state
parameter, $w$~(Holder et al.~2001). In addition, phenomenological
modeling of the dark matter in terms of the halo model (reviewed in
Cooray \& Sheth~2002) requires knowledge of the halo mass distribution
and density profiles, as does the halo occupancy distribution (HOD)
approach to modeling galaxy bias.

Because accurate theoretical results for the mass function (and other
halo properties) do not exist, many numerical studies of halos and
their properties, and of the mass function, have been carried out over
widely separated mass and redshift ranges. Despite the intuitive
simplicity and practical importance of the halo paradigm, halo
definitions and characterizations have been somewhat {\em ad hoc},
mostly because of the lack of an adequate theoretical framework.  For
the purposes of this work, there are two crucial results that have
been well-established by the numerical studies. The first is that
spherically averaged halo profiles are well-described by the
two-parameter NFW profile~(Navarro et al.~1996, 1997) (this shape is
consistent with observational studies of clusters), and second, that a
simple ``universal'' form for the FOF halo mass function (with link
length, $b=0.2$) holds for standard cold dark matter
cosmologies~(Jenkins et al.~2001). A detailed understanding of both of
these numerically established results remains elusive.

\begin{figure}[t]
  \hspace{-1.3cm}\includegraphics[width=120mm]{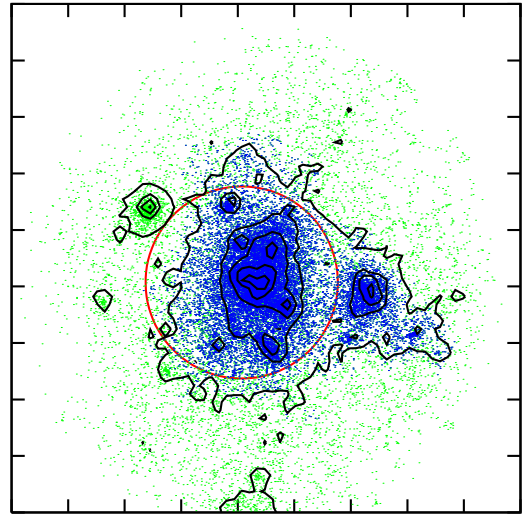} \caption{Different
    halo definitions for the same particle distribution in a
    simulation. The green points show all particles in a sphere
    centered around the minimum potential FOF particle and with radius
    1.1 times the distance to the farthest FOF member ($b=0.2$). The
    black contours are for the two dimensional density field projected
    onto the $z-$direction as calculated from all the particles. The
    blue particles show the actual FOF halo members. The red circle
    shows the SO halo centered around the same point as the FOF
    halo. The box spans approximately 3.15$h^{-1}$Mpc in $x$- and
    $y$-direction, $R_{200}$ is approximately 0.6$h^{-1}$Mpc. The FOF
    mass of the halo is 6.70$\times 10^{13}h^{-1}$M$_\odot$, the SO
    mass of the main halo is 4.91$\times 10^{13}h^{-1}$M$_\odot$ and
    the SO mass of the major subclump on the right (which belongs to
    the FOF halo) is 8.50$\times 10^{12}h^{-1}$M$_\odot$. The small
    subclump on the left (which was neither included in the FOF halo
    nor in the SO halo) is 2.97$\times 10^{12}h^{-1}$M$_\odot$.  This
    plot demonstrates how closely the FOF halo boundary tracks an
    isodensity contour.}
\label{contour}
\end{figure}

The universality of the FOF mass function has been recently verified
to the level of $\lesssim 10\%$ accuracy for essentially all
observationally relevant redshifts ($z \lesssim 10$) by several
simulation efforts (e.g., Heitmann et al.~2006, Reed et
al.~2003,~2007, Luki\'c et al.~2007). The result is potentially very
useful, because at this level of accuracy there is no longer any
reason to simulate individual cosmologies, as the universal form
already covers the parametric region of interest. There is one serious
problem, however: the universal form of the mass function does not
hold for the SO mass as defined and used by observers when determining
the masses of galaxy groups and clusters~(White~2001,
Voit~2005). Unlike the SO criterion, the FOF method (Einasto et
al.~1984, Davis et al.~1985) does not determine a
(spherically-averaged) overdensity structure, but instead defines an
object bound by some isodensity contour (Fig.~\ref{contour}). In
principle, isodensity-based methods can be used in observations, but
require significantly more work than the SO approach.

At this point, one could ask the question whether the SO and FOF
masses could be mapped to each other if more information regarding
halo properties were avalilable. (Or one could forsake universality
and attack the SO mass function problem directly via simulations,
e.g., Evrard et al.~2002, Tinker et al.~2008.) The aim here is to
proceed along the first path and investigate whether an effective
solution to the problem can be found. (For an earlier discussion, see
White~2002, who noted that FOF and SO masses are correlated, but with
a significant scatter.) We first show that even for perfect NFW halos,
there is no simple direct mapping between FOF and SO masses, because
of a significant dependence on the halo concentration. The mapping
depends as well on the number of particles sampling a given halo,
something that needs to be taken into account when interpreting
results from simulations. However, we establish the useful result that
for NFW halos sampled by a given number of particles, a two-parameter
map utilizing concentration and particle number indeed connects the
two masses (with a small Gaussian scatter, quantified below in
Section~3).

The key question is whether these relationships for idealized NFW
halos survive when applied to the more realistic case of halos within
cosmological N-body simulations. We find that this is indeed the case
for halos that can be considered to be relatively isolated (a notion
to be made more concrete in Section~3), and not possess significant
substructure; i.e., approximately $80-85\%$ of all halos in the
mass-range $10^{12.5}-10^{15.5}h^{-1}M_\odot$ explored by the
simulations. (This fraction of isolated halos is close to the
conclusion of Evrard et al.~2008 who anlayzed results from a large
suite of simulations.) For these halos, the two-parameter map derived
above succeeds remarkably well in accurately converting the FOF mass
function to the corresponding SO mass function, at the $\sim 5\%$
level -- the current level of descriptive accuracy as limited by the
robustness of halo definitions and numerical results from simulations
(Luki\'c et al.~2007, Heitmann et al.~2007). 
We show that the concentration dependence of
the FOF-SO mass relation is significant at the current levels of
accuracy for the determination of halo masses.  Conversion between FOF
and SO masses will incur significant error if halo concentration is
not considered.  To transform between the FOF and the SO mass
function, the scatter in concentration must also be considered.  Our
work has implications for observationally determined mass functions,
and for HOD and other methods of deriving mock galaxy catalogs.

An additional point is that, in the N-body simulations, there not only
exists a simple relationship between the halo concentration and the
SO (or FOF) mass with a (relatively) large scatter, but that the
scatter can be very well fit by a Gaussian distribution at a given
mass. Using this simple concentration-mass relation and its Gaussian
variance, one may go directly from the FOF mass function to the SO
mass function or vice-versa. This procedure solves the mass function
mapping problem for the subset of isolated halos, which comprise the
bulk of the halo population. It does not, however, enable one to
transform from the universal FOF mass function to a chosen SO mass
function because of the $15-20\%$ fraction of FOF halos with irregular
morphologies, most of which are ``bridged'' halos (density peaks
connected by high density filaments or ridges). A potential way around
this difficulty is to treat explicitly the ``multiplicity'' of
apparently discrete SO halos within FOF halos in the transformation
between FOF and SO mass functions.  This possibility is under
investigation.

Based on our runs for two cosmologies, we have good evidence that the
fraction of bridged halos rises as a function of mass, and that this
fraction is also ``universal'', i.e., more or less independent of the
cosmology when written in units of $M/M_*$, where $M_*$ is the
characteristic halo mass-scale set by matching the {\em rms} linear
density fluctuation to the threshold density for collapse. We also
find that the fraction of halos with major satellites as a function of
the satellite mass fraction (with respect to the main halo) is
cosmology dependent. This may pave the way for constraining cosmology
from clusters of galaxies in a new way, essentially independent of the
sampling volume, and therefore with enhanced immunity against
selection effects. At the very least, using the major satellite halo
fraction should provide a valuable cross-check for cosmological
constraints derived from the mass function in the conventional manner.

\section{Mass Definitions}

The spherical overdensity and friends-of-friends methods are the two
main approaches to defining halos and their associated masses in
simulations. SO identifies halos by identifying spherical regions with
prescribed spherical overdensities $\Delta$:
\begin{equation}
M_{\Delta} = \frac{4 \pi}{3}R_{\Delta}^3\, \Delta \rho_c \ ,
\end{equation}
where $\rho_c$ is the critical density. (Overdensities are sometimes
stated with respect to the background density: $\rho_b = \Omega_m
\rho_c$, here we restrict ourselves to defining them with respect to
$\rho_c$.)  An often-used value for the overdensity is $\Delta=200$,
roughly the theoretically predicted value given by the spherical
collapse model, $18 \pi^2$, for virialized halos in an Einstein-de
Sitter universe. For the currently favored $\Lambda$CDM model
($\Omega_{\Lambda} = 0.7$, $\Omega_{m} = 0.3$), spherical collapse
actually predicts a smaller overdensity at virialization: $\Delta
\approx 100$. X-ray observers, on the other hand, prefer higher
density contrasts, $\Delta = 500$ or $1000$, because strucutures on
those scales are much brighter, and more relaxed compared to the outer
regions.

\begin{figure*}[t]
  \includegraphics[width=175mm]{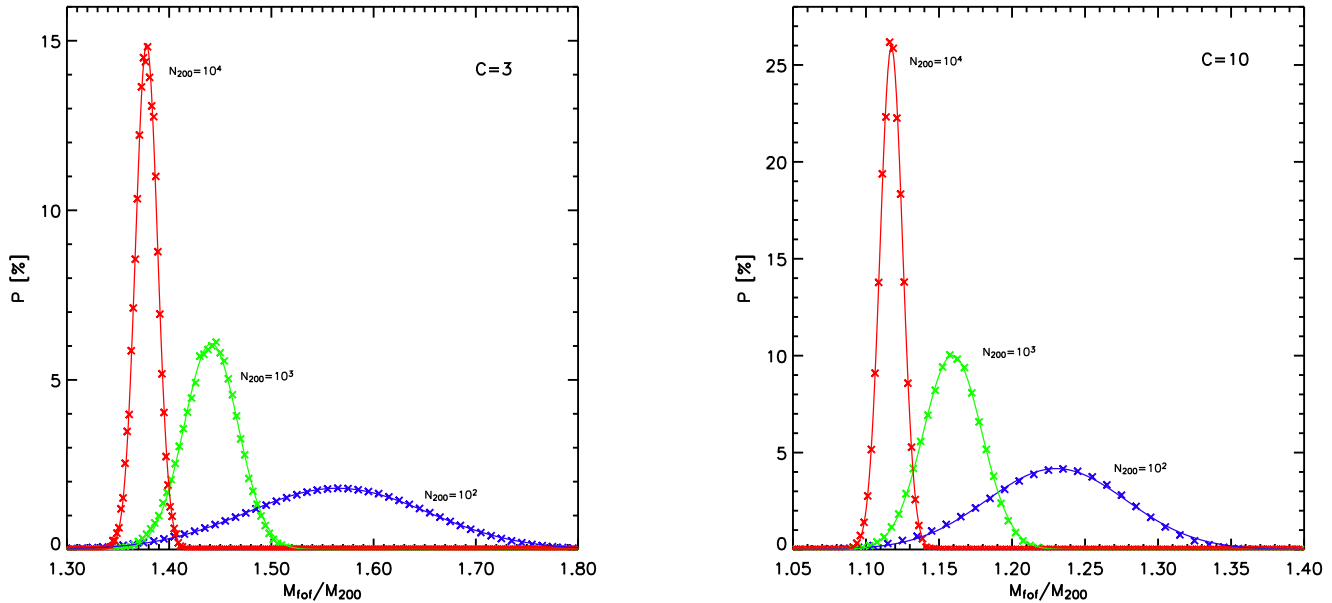}
  \caption{Distribution of $b=0.2$ FOF masses for NFW halos with
    concentrations $c=3$ (left panel), and $c=10$ (right panel),
    sampled with different particle numbers: 100 (blue), 1000 (green),
    10000 (red). The number of Monte Carlo samples are $10^6$, $10^5$,
    and $10^4$ for $N_{200}=100$, 1000, and 10000, respectively. The
    solid curves are Gaussian fits. Note that the two panels have
    different units along both axes.}
\label{mockdev}
\end{figure*}

The main drawback of the SO mass definition is that it is somewhat
artificial, enforcing spherical symmetry on all objects, while in
reality halos often have an irregular structure (e.g., White~2002).
For some applications, such an approach may be well founded
(e.g. X-ray cluster analysis for relaxed clusters), but may not be
universally applicable.  Furthermore, defining an SO mass can be
ambiguous, since for two close density peaks, the corresponding SO
spheres might overlap, and one has to decide how to distribute
particles between them (or assign them to both, breaking mass
conservation).

The FOF algorithm, on the other hand, is not based on the notion of a
certain overdensity structure, but defines instead an object bound by
some isodensity contour. The mass of a halo is then simply the sum of
all particles inside a given contour. By linking particles which are
separated at most by the distance $ll = b n^{-1/3}$ (where $n$ is the
number density of particles in the simulation, and $b$ is the
so-called ``linking length''), the FOF method, in effect locates an
isodensity surface of
\begin{equation}
\rho_{iso} \approx k b^{-3} \rho_b\ ,
\label{eqn:iso}
\end{equation}
where $k$ is a constant of order 2~\cite[]{FRENK1988}. For $b=0.2$,
and the concordance $\Lambda$CDM cosmology, this leads to $\rho_{iso}
= 75 \rho_c$. Given their percolation-centric nature FOF halos can
have complicated shapes and topologies (Fig.~1).

\section{Mass Mapping from Mock Halos}

In order to address the relation of FOF and SO masses, we first turn
to a controlled test using idealized ``mock'' halos. These are
taken to be spherical dark matter halos with the NFW density profile:
\begin{equation}
\rho(r) = \frac{\rho_s}{r/r_s\left( 1+ r/r_s\right)^2}\ ,
\end{equation}
where $\rho_s$ and $r_s$ are the core density and scale radius
respectively.  Instead of $\rho_s$ and $r_s$, it is often convenient
to use physically more transparent quantities: the SO mass
$M_{\Delta}$ and the concentration $c=r_s/R_{\Delta}$:
\begin{equation}
\label{rho_s}
\rho_s = \frac{\Delta \ \rho_c \ c^3}
{3 \ \left [ \ln (1+c) - c/(1+c) \right] }\ ;  
\end{equation}
\begin{equation} 
\label{r_s}
r_s = \frac{1}{c} \left[ \frac{3\ M_{\Delta}}{4\ \pi \ \Delta \ \rho_c}
\right]^{1/3}\ .
\end{equation}
The cumulative mass within a radius $r$ can be calculated as:
\begin{eqnarray}
M(r) &=& \int_0^r 4 \pi r^2 \frac{\rho_s}{r/r_s\left( 1+
    r/r_s\right)^2}  dr    \nonumber \\
     &=& 4 \pi \rho_s r_s^3 \left [ \ln (1+r/r_s) - (r/r_s)/(1+r/r_s)
     \right]\ . 
\label{eqn:cumulative}
\end{eqnarray}
While it is still unclear whether the very inner parts of the halos
($\sim$ 1\% of $R_{200}$) have density profiles steeper than NFW
(e.g., Ghigna et al.~2000, Jing \& Suto~2000, Klypin et al.~2001,
Navarro et al.~2004, Reed~et~al.~2005), the inner asymptotic slope is
not of concern here, and does not affect our results.

We generate mock NFW halos in the following way: first we fix the SO
mass ($M_{\Delta} = M_{200}$) of a halo and choose the number of
particles which will reside in it ($N_{200}$). We then populate the
halo with particles according to the NFW distribution such that we
enforce the desired mass to be $M_{200}$ within the radius
$R_{200}$. We then extend the NFW distribution further out -- adding
particles to a ``halo tail''. The choice of $\Delta=200$ can easily be
changed to some other desired value such as $\Delta=500$ or 1000 as
more appropriate for cluster studies. In any case, for a given NFW
profile choice, all overdensity masses are immediately fixed, so there
is no lack of generality in our specific choice (which corresponds to
an approximate notion of the ``virial mass'', Navarro et al., 1996,
1997).

\begin{figure}[t]
  \includegraphics[width=80mm]{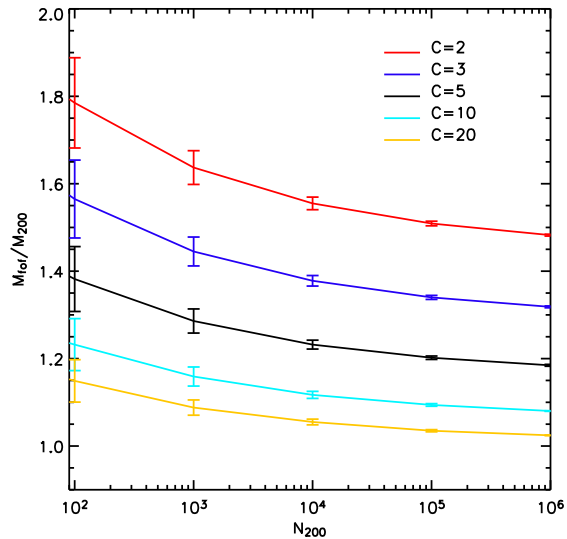}
  \caption{Ratio of the ($b=0.2$) FOF mass to M$_{200}$ for NFW mock
    halos with different concentrations and particle number, $N$, but
    the same value of $M_{200}$. Low concentration halos have up to a
    factor of two higher FOF mass than M$_{200}$. For high
    concentration halos, the ratio of the two mass definitions is
    closer to unity, the FOF mass being always higher.}
\label{mockhalos}
\end{figure}

Having fixed $M_{200}$ for all the mock halos, we now determine the
FOF mass for every halo. Because the particles are randomly sampled
inside a halo (following the NFW density profile), one cannot expect
that for every realization of a mock halo, the FOF finder will return
exactly the same mass. Given a large number of mock halos with the
same density profile and statistical independence of the realizations,
the central limit theorem predicts a Gaussian distribution for the FOF
masses. Indeed, just as expected, a normal distribution gives an
excellent description for $M_{FOF}/M_{200}$.  Thus, one can not only
determine to what SO mass a certain $M_{FOF}$ corresponds (on
average), but can also quantify the systematic deviation of an FOF
halo finder through a standard deviation (Figs.~\ref{mockdev}). 
The Gaussian spread of FOF masses is centered around
a mean value that shifts systematically with the number of sampling
particles, $N$, as empirically noted by Warren et al.~(2006)
(Fig.~\ref{mockhalos}).

Besides this $N$-dependence, we also wish to examine how
$M_{FOF}/M_{\Delta}$ depends on the underlying profile. We have found
that this dependence leads to another source of bias for FOF masses
relative to SO masses.  In Fig.~\ref{mockhalos}, we show average
values of $M_{FOF}$ for a range of particle numbers and
concentrations.  It is clear that one cannot accurately match a given
$M_{200}$ to a corresponding $M_{FOF}$ without the concentration being
specified. Concentration variation from $c\sim 20$ (typical for
galaxies) to $c\sim 5$ (typical for clusters) (Bullock et al. 2001;
Eke, Navarro, \& Steinmetz 2001) corresponds to systematic FOF mass
shifts of $\sim 30\%$, much larger than can be tolerated by the
accuracy to which the FOF mass function can currently be determined
numerically ($\sim 5\%$). For any given $N_{200}$, this concentration
dependence follows the functional form:
\begin{equation}
\frac{M_{FOF}}{M_{200}} = \frac{a_1}{c^2} + \frac{a_2}{c} + a_3 \ ,
\label{mock_fit}
\end{equation}
where the coefficients $a_1$, $a_2$, $a_3$, depend on $N_{200}$ only
(Table~\ref{coeffs}).

\begin{table*}
\begin{center} 
\caption{\label{coeffs} Best Fit Coefficients}
\begin{tabular}{c|cccccccc}
\tableline\tableline
  & & & & &  $N_{200}$ & & \\
\raisebox{1.4ex}[0pt]{Coeff.} & 100 & 600 & $10^3$ & $3 \times 10^3$ &
$6 \times 10^3$ & $10^4$ & $10^5$ &
$10^6$ \\
\hline
 $a_1$ & -0.3887  & -0.3063  & -0.2790  & -0.2368 & -0.2210
& -0.1970  & -0.1642 & -0.1374\\
 $a_2$ & 1.6195  & 1.4130  & 1.3669  & 1.2849 & 1.2459
& 1.2157  & 1.1392 & 1.0900\\
 $a_3$ &  1.0715  & 1.0313  & 1.0226 & 1.0081 & 1.0008
& 0.9960 &  0.9800 & 0.9714\\
\tableline\tableline

\vspace{-1.5cm}

\tablecomments{Best fit coefficients for different $N_{200}$, as obtained 
               from the mock halo analysis. For all values of
               $N_{200}$, the functional form of the fit is 
               given by Eqn.~(\ref{mock_fit}).}
\end{tabular}
\end{center}
\end{table*}

Well-sampled halos, with $N >1000$, are characterized by a small
variance in the $M_{FOF}/M_{200}$ ratio, with a maximum value of
$\sigma\sim 0.02-0.03$, depending on the concentration. With such a
low intrinsic scatter in the mass relationship for a given
concentration, the logical next step is to see whether the mean
$M_{FOF}(M_{200},c)$ relationship obtained from the mock NFW halos
actually applies to individual halos in N-body simulations. Here, it
should be noted that actual simulated halos are not expected to be
spherical due to the episodic and anisotropic nature of mass
accretion, and in fact are much better described as ellipsoids (Kasun
\& Evrard~2005, Allgood et al.~2006). Nevertheless, as we are
interested in an averaged quantity, the halo mass, an approach based
on idealized halos may well provide an adequate description. This
expectation turns out to be valid, as shown below.

\section{Mass Mapping in N-Body Simulations}

In order to investigate the validity of the mock halo mass
relationships, we use results from four cosmological simulations for
two flat $\Lambda$CDM cosmologies, each simulated with 174 and 512
$h^{-1}$Mpc boxes. The pre-WMAP, high-$\sigma_{\rm 8}$ cosmology has
the following parameters: matter density, $\Omega_m=0.3$; dark energy
density, $\Omega_\Lambda=0.7$; fluctuation amplitude, $\sigma_{\rm
  8}=1.0$; Hubble constant $h=0.7$ (in units of 100 km s$^{-1}$
Mpc$^{-1}$); primordial spectral index, $n_s=1$; and the Bardeen et
al.~(1986) transfer function with $\gamma=\Omega_m h$.  For the WMAP~3
compatible cosmology runs, the parameters are: $\Omega_m=0.26$,
$\Omega_\Lambda=0.74$, $\sigma_{\rm 8}=0.75$, $h=0.71$, $n_s=0.938$,
and a transfer function generated using {\small CMBFAST} (Seljak \&
Zaldarriaga~1996).  We use the parallel gravity solver {\small
  GADGET2}~\cite[]{gadget2} to follow the evolution of $512^{3}$ dark
matter particles starting from a redshift $z=99$, high enough to satisfy
the initial redshift requirements given in Luki\'c et al.~2007.  The
particle masses are $3.3 \times 10^{9}$ and $8.3 \times10^{10} h^{-1}
M_\odot$ for the high-$\sigma_{\rm 8}$ run, and $2.8 \times 10^{9}$
and $7.2 \times10^{10} h^{-1} M_\odot$ for the WMAP~3 cosmology.
These masses are small enough to comfortably resolve groups and
clusters to the level required for this study (see e.g. Power et
al.~2003, Reed et al.~2005, Neto et al~2007). The FOF mass functions
from these simulations are in very close agreement with the results of
Luki\'c et al.~(2007), well within a few percent.  By using
cosmologies with normalizations that bracket the currently favored
cosmology (e.g., Spergel~et al.~2007), we are able to show that our
results are applicable to any likely cosmology, once (cosmology
dependent) halo concentrations are specified.

To carry out a realistic test of the mass relationships, we adopt the
following procedure: (i) First run an FOF halo finder on the final
particle distribution, and define halo centers by identifying the
local potential minima, for all halos with $N>1000$. (ii) Construct
individual SO profiles around these minima, thereby determining
$M_{200}$.  The halo density is computed in 32 logarithmically
equidistant bins, and we fit the NFW profile treating both $r_s$ and
$\rho_s$ as free parameters. As a consistency check, we use an
alternative approach, where $M_{200}$ is measured directly from the
mass within a sphere, and NFW is treated as a one-parameter function
(by fixing $\rho_s$ such that the enclosed overdensity is
$200\rho_c$). No significant differences were found between the two
approaches.

\begin{figure}[t]
  \includegraphics[width=80mm]{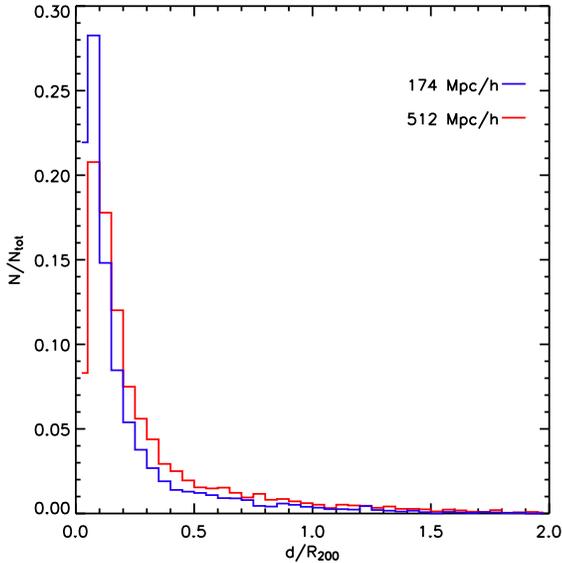}
  \caption{Distribution of distances between FOF center of mass, and
    potential minimum for 512 $h^{-1}$Mpc box (red) and 174 $h^{-1}$Mpc box
    (blue), scaled by R$_{200}$}
\label{halo_centers}
\end{figure}

\begin{figure}[t]
  \includegraphics[width=80mm]{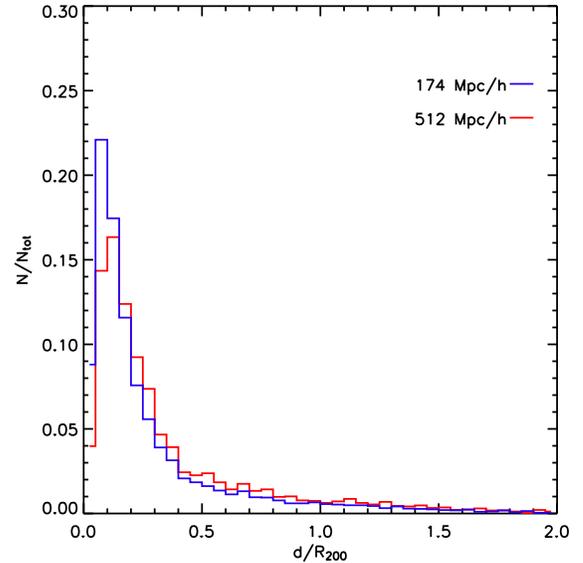}
  \caption{Same as Fig.~\ref{halo_centers}, but for the WMAP~3
    cosmology.} 
\label{halo_centers_wmap}
\end{figure}

The $N>1000$ halo particle cut keeps the variance in the mass ratios
small (Cf. Figs.~\ref{mockdev}-\ref{mockhalos}) and also allows stable
calculations of the individual halo concentrations. [Details of the
procedures followed will be given elsewhere (Reed~et al., in
preparation).] For each FOF halo we find its center of mass from
all the particles linked together by the halo finder.  On occasion,
the FOF finder connects apparently distinct halos (bridging); these
halos may well be in some stage of merging. Since it makes little
sense to define an SO profile and an associated concentration for
very close halos and those undergoing major mergers, we use the
distance between the center of mass and the potential minima to
exclude such halos. In Figs.~\ref{halo_centers} and
\ref{halo_centers_wmap}, we show the distribution of that distance
($d$) for all halos with $N > 1000$ from both of the simulation
boxes. While most of the halos appear to be isolated objects where the
difference between the two center definitions is due to substructure,
there are outliers at high mass, and even objects where the FOF center
of mass is more than $R_{200}$ away from the potential minimum!

\begin{figure}[t]
  \includegraphics[width=80mm]{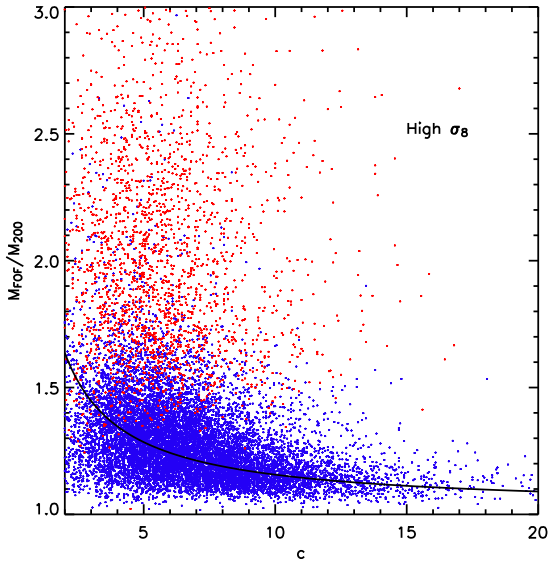}
  \caption{Scatterplot of the ratio of FOF and SO(200) masses from the
    simulations as a function of the measured concentration for (i)
    halos passing the criterion $d/R_{200}<0.4$ (blue), where $d$ is the
    distance between the center of mass and the potential minima (see
    discussion in the text), and (ii) halos not passing this criterion
    (red). The solid line shows the mock halo prediction for halos
    with particle number, N$_{200}=10^3$, which dominate the sample.}
\label{banana}
\end{figure}

\begin{figure}[t]
  \includegraphics[width=80mm]{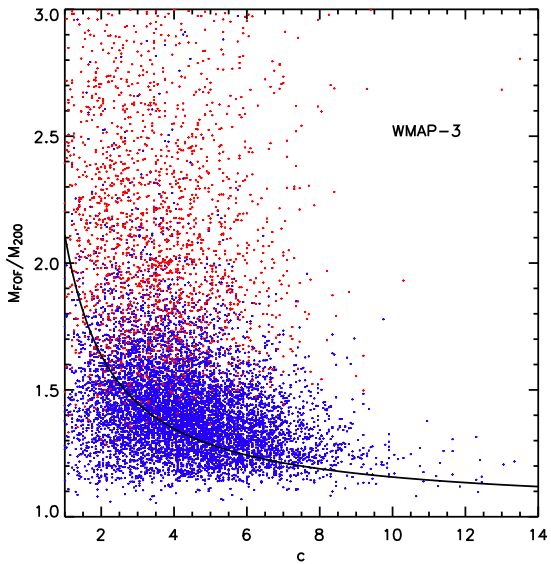}
  \caption{Same as Fig.~\ref{banana}, but for the WMAP 3 cosmology.
    Note that the x-axis has a different scale.}
\label{banana_wmap}
\end{figure}

To proceed further, we first set aside all halos with
$d/R_{200}>0.4$. Although this cut is somewhat arbitrary, the results
are relatively insensitive to the particular choice, as discussed
below. Furthermore, the mock halo analysis on regular NFW halos shows
that, even at low concentrations, one expects approximately
$M_{FOF}/M_{200}\sim 1.5$ (Cf. Fig.~\ref{mockhalos}). Larger values
therefore are a signal of a potential merger, as was verified directly
by confirming with the simulation results. In Figs.~\ref{banana} and
\ref{banana_wmap}, where we plot both ``isolated'' (blue) and
``bridged'' (red) halos, the strong correlation between our cut, based
on the difference between halo mass and potential centers, and the
high values of $M_{FOF}/M_{200}$ (with respect to the mock halo
expectation) can be easliy verified.

Finally, we compare our cutoff with an SO analysis of FOF halos: for
each halo we find the minimum potential particle and $R_{200}$ around
it, and than move to the next particle in the potential hierarchy
which resides outside $R_{200}$ (if inside, we define it as a piece of
substructure rather than a `satellite halo' bridged by the FOF
procedure), find $R_{200}$ and $M_{200}$ for the second halo, and
iterate this procedure until all FOF particles are exhausted. When
separate SO halos overlap we assign particles in the overlapping
region to all SO halos, keeping the overdensity idea straightforward,
but breaking mass conservation. Of course, if one goes down to a few
particles, then virtually all FOF halos will be resolved into multiple
SO objects. But if the threshold of the satellite mass is raised to
20\% of the main halo mass, most of the FOF halos appear as a single
SO halo.  The two methods: $d/R_{200}>0.4$, and
$M_{satellite}/M_{main} > 0.2$ correlate extremely well, agreeing in
85-90\% of all cases (the agreement is worse for larger masses, and
better for smaller halo masses). This gives us additional confidence
that our cutoff criterion separates isolated from bridged halos.  We
will return to an analysis of the excluded halos (by both of the
discussed exclusion criteria) in Section~\ref{bridged}.

\begin{figure}[t]
  \includegraphics[width=80mm]{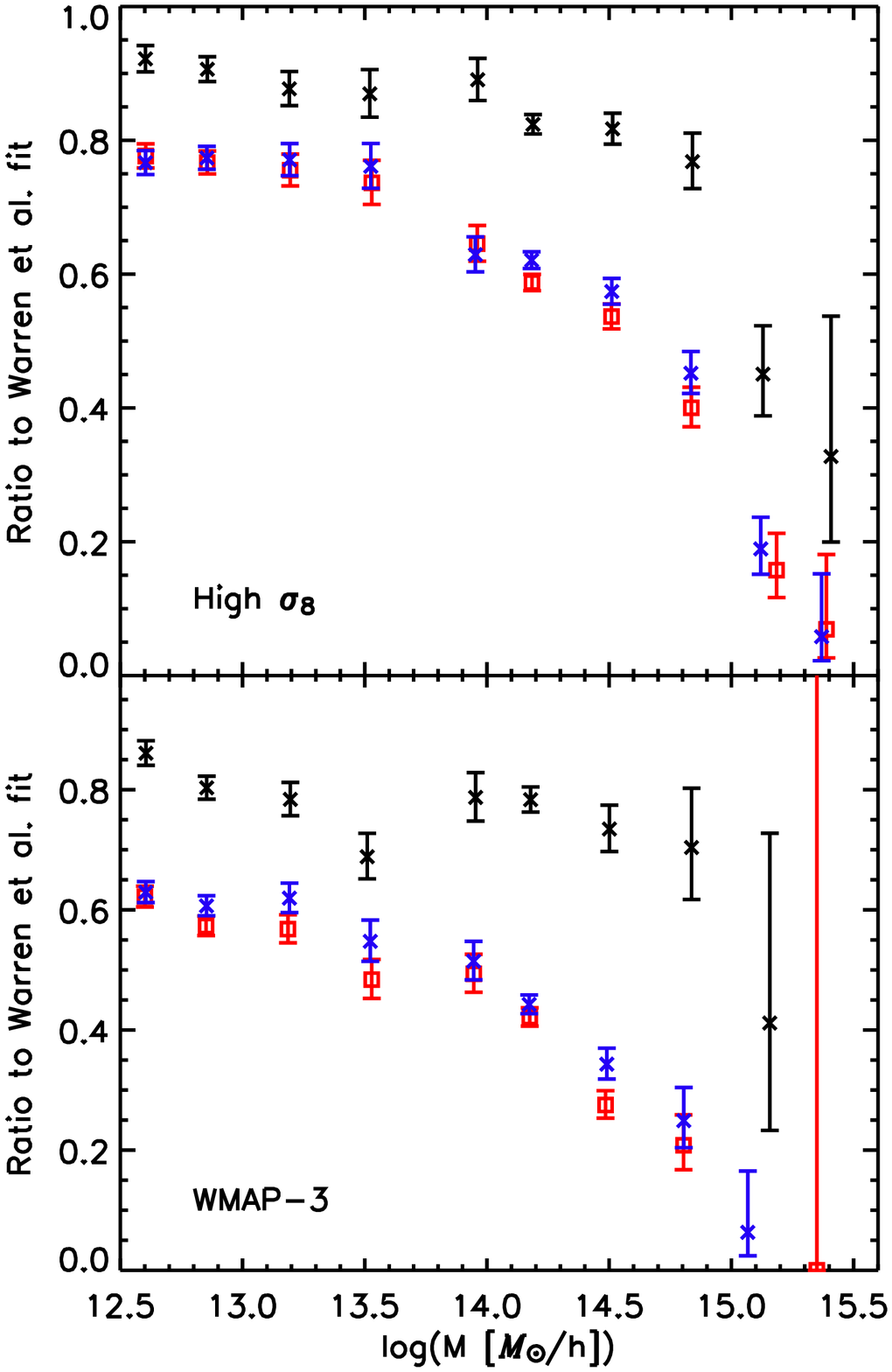}
  \caption{Measured mass functions normalized to the Warren~et
    al.~(2006) fit as an (arbitrary) reference, for High $\sigma_{\rm 8}$
    (upper panel) and WMAP-3 cosmology (lower panel).  
    Black: FOF halo masses
    with $b=0.2$ and bridged halos removed as shown in
    Fig.~\ref{banana}. Red: $M_{200}$ masses measured from the
    simulation for the same set of halos, and using the same (FOF)
    halo centers. Blue: The mass function for $M_{200}$ halos using
    the idealized mock halo prediction (Fig.~\ref{mockhalos} and
    Table.~\ref{coeffs}), the measured FOF masses for each halo as mapped to
    the predicted SO mass. The agreement between measured (red) and
    predicted (blue) mass functions is excellent, better than $5\%$.}
\label{halo_halo}
\end{figure}

\begin{figure}[t]
  \includegraphics[width=80mm]{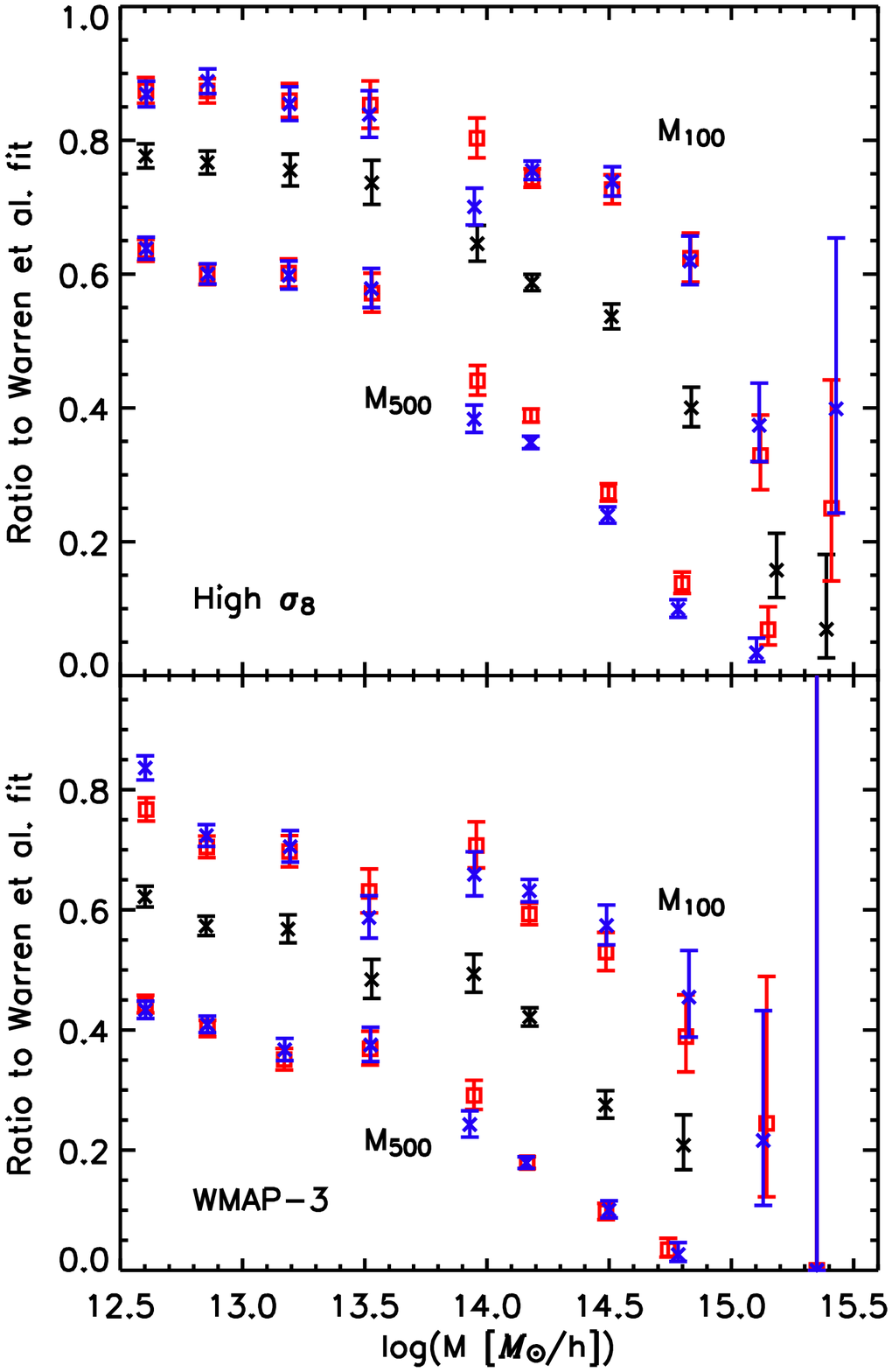}
  \caption{Testing the mapping to masses other than $M_{200}$ with
    mass functions shown as in Fig.~\ref{halo_halo}, 
    for High $\sigma_{\rm 8}$
    (upper panel) and WMAP-3 cosmology (lower panel).
    Black: $M_{200}$
    masses measured from the simulation. Red: $M_{100}$ and $M_{500}$
    masses measured from the simulation using the same halo
    centers. Blue: Idealized NFW predictions for $M_{100}$ and
    $M_{500}$ using the measured $M_{200}$ mass for each
    halo. Measured and predicted quantities (red vs. blue) are again
    in very good agreement.}
\label{so_so}
\end{figure}

The halo exclusion cut eliminates only about 15-20\% of all halos, so
it is not very statistically significant, though certainly not
negligible. For the retained halos, we now apply the
$M_{FOF}(M_{200},c)$ relationship determined by the mock halo results
of Fig.~\ref{mockhalos}, as encapsulated in the fits specified in
Table~\ref{coeffs}. The results of this halo by halo mass mapping test
are shown in Figs.~\ref{halo_halo} and \ref{so_so} for the
mass function, where the measured mass functions are displayed in
terms of a ratio to a fitting form for the FOF mass function given by
Warren~et al.~(2006). (This ratio is taken only for ease of
interpretation, as any other mass function fit would have done just as
well.)  The undernormalization of the FOF mass function relative to
the fit is simply due to the exclusion procedure described above. Note
that the FOF and SO mass functions, as numerically determined, differ
by as much as $20-40\%$ depending on the mass bin. However,
application of the mock halo mass relationship to every individual FOF
halo {\em correctly reproduces the SO mass function at the $5\%$
  level}, the current (numerical) limiting accuracy of mass function
determination. The success of this simple mapping idea is a testimony
to the accuracy of the NFW description for (spherically averaged)
realistic halos in simulations, and consistent with the overall
conclusion of Evrard et al.~2008, that the vast majority of
cluster-scale halos are structurally regular.

Using the expression for the cumulative NFW mass
[Eqn.~(\ref{eqn:cumulative})], we can find the mass for any desired
overdensity $\Delta$ in terms of $M_{200}$; defining $M_c =
M_{\Delta}/M_{200}$, we have:
\begin{equation}
M_c = A(c) \left[ \ln \left( 1 + \sqrt[3]{\frac{200}{\Delta}} M_c c \right)
            - \frac{\sqrt[3]{\frac{200}{\Delta}} M_c c}
              {1+\sqrt[3]{\frac{200}{\Delta}} M_c c} \right]\ ,
\label{nfw_transform}
\end{equation}
where $A(c)$ is a prefactor which depends on c only:
\begin{equation}
A(c) = \frac{1}{\ln (1+c) - c/(1+c)}\ .
\end{equation}
Employing this approach one can easily move from one SO mass function
to another, and in Fig. \ref{so_so} we show that
this mass transformation gives accurate results for halos in
simulations. Furthermore, this shows that if one is interested in any
overdensity other than 200 (as considered in our mock halo analysis),
our best fit for $M_{FOF}/M_{200}$ [Eqn.~(\ref{mock_fit}) and Table
\ref{coeffs}] can simply be rescaled for any $M_{\Delta}$ using
Eqn.~(\ref{nfw_transform}).

The results shown in Figs.~\ref{halo_halo} and \ref{so_so}
depend only weakly on the cut imposed by a particular value of
$d/R_{200}$. Choosing a value below $d/R_{200}=0.4$ such as 0.3 is
more conservative; one loses more halos (another $5\%$), but the mass
function mapping results remain excellent. Increasing the cut
threshold to 0.5 adds $5\%$ more halos while the mapping accuracy
remains more or less the same. Beyond this point the results slowly
degrade, as is to be expected.

With this important result at the level of individual halos in hand,
the global mass function can be realized without knowing individual
halo concentrations, and independent of cosmology, provided one has a
form for the (mean) concentration-mass relation for SO (or FOF) halos
as well as the PDF for the scatter in this relation. The latter
cannnot be ignored since the scatter in the concentration-mass
relation is known to be significant (Jing~2000, Bullock et al.~2001,
Eke et al.~2001, Macci{\`o} et al.~2007, Neto et al.~2007). In the
mass regime typical for clusters, i.e., halo masses above $\sim
3\times 10^{14}$ $h^{-1}M_\odot$, the variation in concentration with mass
is in fact much smaller than the concentration scatter for halos of
similar mass.  We have carried out several detailed simulations, aside
from the ones mentioned here, to establish the cosmology dependence of
the concentration-mass relation, $c(M_{200})$, and its associated
scatter, $\sigma_c(M_{200})$ [or $\sigma_c(M_{FOF})$] (Reed~et al., in
preparation), which provides all the required information for mapping
mass functions.  The scatter is very well described by a Gaussian PDF
at each mass bin (for both SO and FOF masses) and has little
variation over the limited mass range relevant for clusters.

\section{The Bridged Halos}
\label{bridged}

\begin{figure*}[t!]
  \includegraphics[width=160mm,height=200mm]{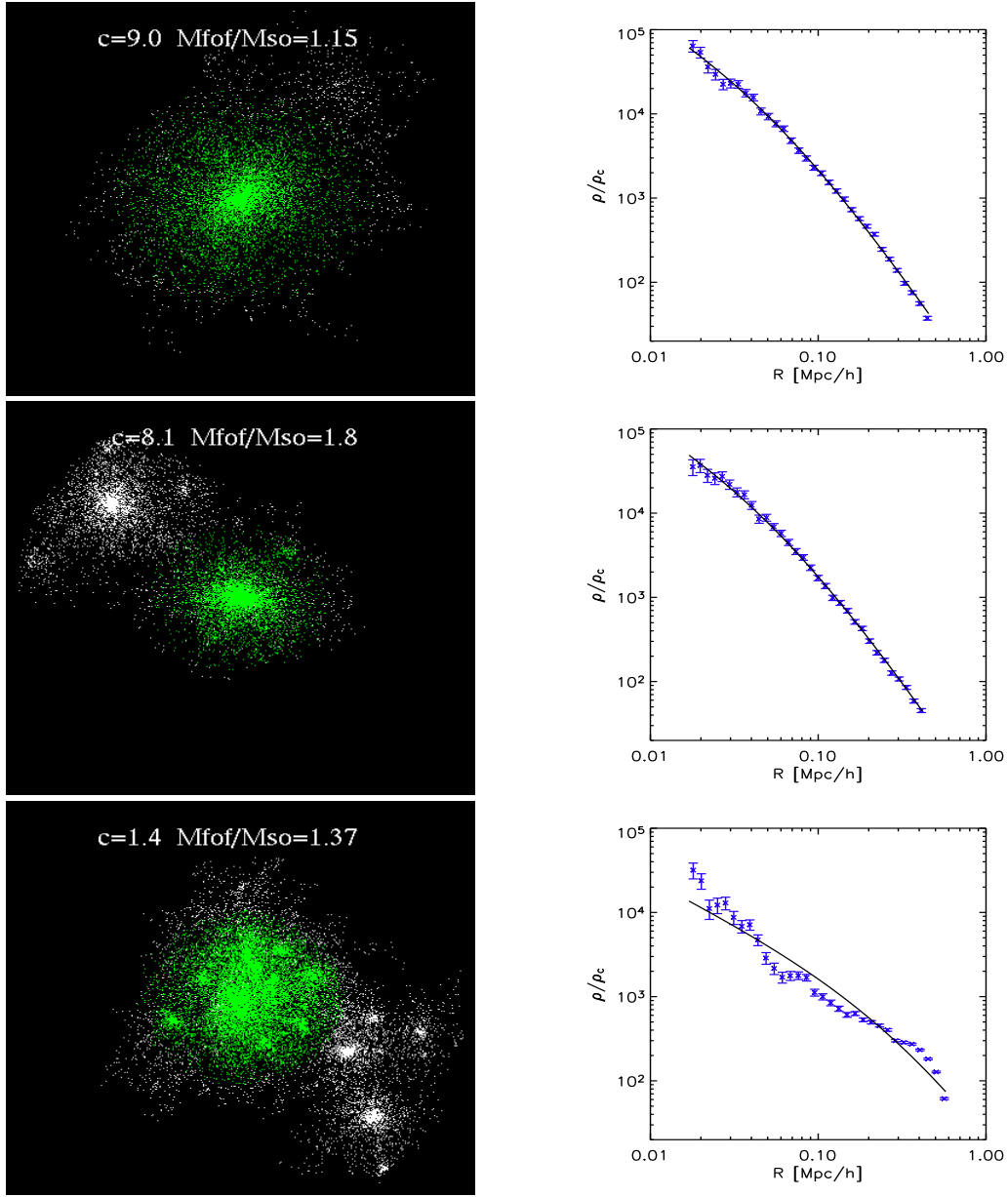}
  \caption{Top panel: A typical isolated FOF halo (FOF-linked
    particles shown as white dots) with NFW concentration, $c=9.0$,
    and $M_{FOF}/M_{200}=1.15$ (profile fit to the right). Green dots
    are particles within $R_{200}$ of the corresponding SO
    halo. Middle panel: An example of a bridged halo. The SO halo
    found at the FOF center has concentration $c=8.1$ (the NFW profile
    fit is a good fit), however the mass ratio $M_{FOF}/M_{200}=1.8$
    is high due to the bridged minor halo in the left upper
    corner. Bottom panel: A halo with major substucture, for which the
    NFW profile is not a good fit.}
\label{halo_images}
\end{figure*}

We now turn to understanding the FOF halos that cannot be simply
mapped as individual NFW profiles. Broadly speaking, we find that
these halos are of two types: (i) Halos with density bridges across
major substructures, and (ii) halos with complex substructure
(``unrelaxed''). Halos of the first type are the ones largely excluded
by our halo mass and potential centers-based cut and correspond mostly
to the high mass-ratio region in Figs.~\ref{banana} and
\ref{banana_wmap}. While our cut is very efficient in terms of
identifying bridged halos, there is a very small contamination
fraction due to chance symmetric bridging which does not lead to
significant differences between the mass and potential minima.  The
second type of halos corresponds largely to the low concentration/low
mass ratio region. Representative halo types are shown in
Fig.~\ref{halo_images}: typical isolated halo (upper panel), bridged
halo (middle panel), and complex substructure (lower panel).

It is clear that the idea of a single concentration or a simple mass
ratio $M_{FOF}/M_{SO}$ makes little sense for either the bridged
halos or the unrelaxed halos. For the unrelaxed halos, absent a
sub-halo analysis, it is not even clear what an appropriate $M_{SO}$
might be. Nevertheless, our exclusion was designed mostly to eliminate
the bridged halos; our results show that the unrelaxed population is
apparently subdominant at least in terms of biasing the mass function
results. Even so, it is clear that the existence of these types of
substructured halos has ramifications for the simple HOD program,
although the quantitative impact needs to be studied.

\begin{figure}[t]
  \includegraphics[width=80mm]{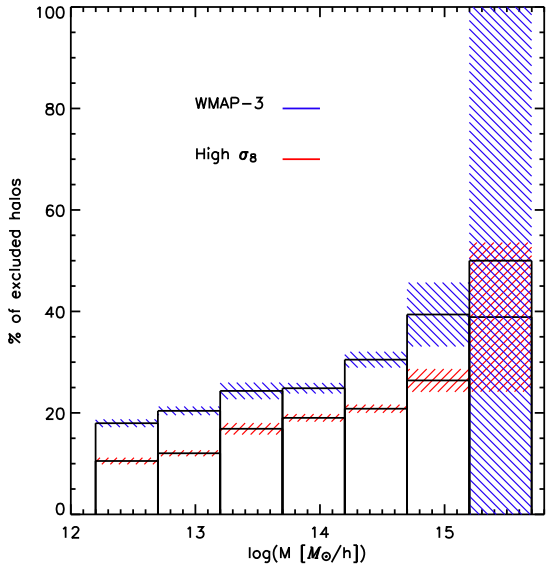}
  \caption{Distribution of bridged halos as a function of mass for the
    high $\sigma_8$ and $WMAP~3$ cosmologies. In both cosmologies, the
    relative fraction of such halos tends to increase with increasing
    mass. The shaded regions are Poisson error bars.}
\label{train_wreck}
\end{figure}

The halos that are bridged by the FOF procedure are typically close
neighbors, the majority being partners in the hierarchical process of
structure formation via halo merging (Busha et al.~2005). Some of
these close neighbors might be ``backsplash halos'' that have
previously been within $R_{200}$ (see Gill, Knebe, \& Gibson 2005;
Ludlow et al. 2008). In both the high-$\sigma_{8}$ and WMAP~3
cosmologies, we find that the fraction of bridged halos has a tendency
to increase with increase in mass. This is as expected from the
hierarchical merging picture since very massive halos are still
forming at the current epoch. This effect is clearly shown in
Fig.~\ref{train_wreck}. We have checked that the two different-sized
boxes (for each cosmology) agree well in the region of overlap,
supporting the argument that numerical effects (finite mass and force
resolution) are negligible for this consideration. (For the two box
sizes, the mass resolution differs by a factor of approximately 25,
and the force resolution by a factor of 3.)

The overall effect can certainly depend on cosmology: the results from
the WMAP~3 simulation are clearly separated from the high $\sigma_{\rm 8}$
cosmology (Fig.~\ref{train_wreck}).  Since the structures grow
differently in the two different cosmologies (due to different
$\sigma_{\rm 8}$ and $\Omega_m$), we can try to parametrize our
exclusion as a function of $M/M_*$, where $M_*$ is the characteristic
collapse mass at the current epoch, defined through:
\begin{equation}
\sigma[M_*(z)] = 1.686 \ ,
\end{equation}
where $\sigma$ is the variance of the linear density fluctuation field
$P(k)$, smoothed by a top-hat filter $W(k,M)$ on a scale $M$, and
normalized to the present epoch $z=0$ by the growth function $d(z)$:
\begin{equation}
\sigma^2(M,z) =
\frac{d^2(z)}{2\pi^2}\int^{\infty}_{0}k^2P(k)W^2(k,M)dk \ .
\label{sig}
\end{equation}

\begin{figure}[t]
  \includegraphics[width=80mm,angle=0]{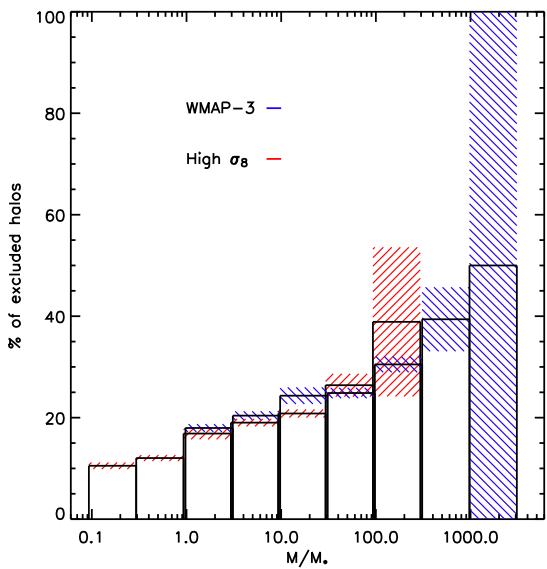}
  \caption{Possible universality of the bridged halo fraction: The
    same data as in Fig.~\ref{train_wreck}, but with the mass now
    scaled by $M_*$.}
\label{train_wreck2}
\end{figure}

As shown in Fig.~\ref{train_wreck2}, with the mass rescaled in terms
of $M_*$, the fraction of bridged halos agrees for the two cosmologies
and may very well be ``universal''. This intriguing fact indicates,
first, that our method of excising bridged halos (the principle, not
necessarily the specific choice of $d/R_{200}>0.4$) is physically
well-motivated.  Second, if the universality is borne out, the bridged
halo fraction can be combined with the cosmology independent mock halo
analysis, to yield a method for translating the universal FOF mass
function to any desired SO mass function. Moreover, these results
suggest that the bridged halo fraction can also provide a separate
probe of cosmology, being particularly sensitive to the same
parameters as the mass function itself (Fig.~\ref{train_wreck}).

\begin{figure}[t]
  \includegraphics[width=80mm,angle=0]{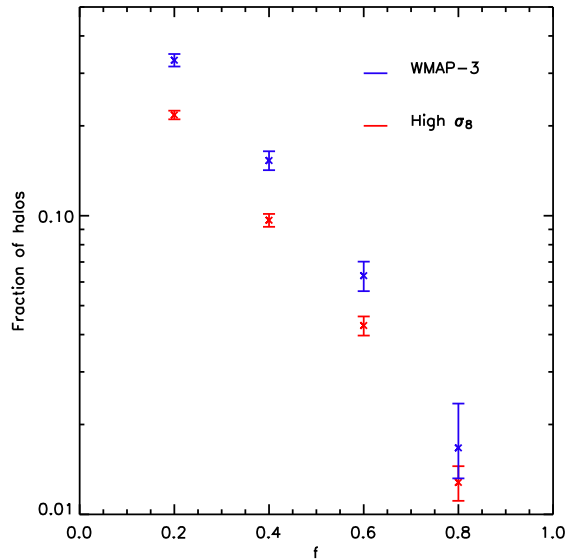}
  \caption{Fraction of the total number of halos in the mass range
    relevant to clusters, ($M_{200} \ge 10^{14} M_{\sun}/h$), as a
    function of the halo satellite mass fraction.} 
\label{satellites}
\end{figure}

An additional way to probe the growth of structure in the Universe
using clusters, aside from the mass function, would be to measure the
fraction of isolated clusters versus those that have (major)
satellites. In our simulations, we measure the fraction of multiple SO
dark matter halos in the mass range of interest for clusters: $M_{200}
\ge 10^{14} M_{\sun}/h$ (see also Evrard et al. 2008). If we plot this
fraction as a function of $f$, where $f$ is defined through
$M_{satellite} \ge f M_{main}$ we find again that the two cosmologies
considered are clearly separated, as shown in
Fig.~\ref{satellites}. The advantage of this analysis compared to the
mass function method is that it does not require measurements in a
controlled volume, and will work for a random sample of observed
galaxy clusters. Depending on observational possibilities (McMillan et
al.~1989, Mohr et al.~1995, Zabludoff \& Zaritsky~1995, Jones \&
Forman~1999, Kolokotronis et al.~2001, Jeltema et al.~2005, Ramella et
al.~2007), this might provide a new way of characterizing cosmologies
using clusters of galaxies, or at least be a valuable method to
cross-check results from mass function constraints.

The halo outliers with values of $M_{FOF}/M_{200}>1.5$ are also a
possible source of systematic bias for certain HOD applications. Given
some halo mass bin above the fiducial mass cutoff for a given HOD, a
bridged halo would be assigned a central galaxy with the same
probability as an isolated halo. The probability of a satellite galaxy
in a bridged halo [with the main halo having high mass companion(s)]
is likely significantly higher than in an isolated halo. Therefore,
applying the same HOD to both halo types would downweight the number
of satellite galaxies, the precise amount depending on the mass range
considered.

\section{Conclusions and Discussion}

We have presented results from an analysis of idealized NFW halos and
N-body simulations with the aim of clarifying the connection between
FOF and SO halos, focusing mainly on the issue of halo masses and
attempting to account for some of the unavoidable difficulties in
simplifying a multi-scale problem in terms of primitive halo
concepts. We found that a large fraction of FOF halos in N-body
simulations ($80-85\%$) are relatively isolated and well-fitted by NFW
profiles. This allows them to have SO counterparts, albeit the mass
mapping is a two-parameter function $M_{SO}=M_{SO}(M_{FOF},c)$,
inferred from the properties of idealized NFW halos ($c$ is the NFW
halo concentration). In principle, this mock halo technique can be
trivially extended to $M_{\Delta}$ with $\Delta$ values more directly
useful for cluster analyses (e.g., $\Delta=500, 800, 1000$), or indeed
to any other useful definition of the observable mass.

The rest of the halos, a fraction of $15-20\%$, appear to be
dominated mainly by bridged halos. These halos consist of apparently
localized structures (visually, or according to the SO halo
definition) linked via density ``ridges'' into a common FOF halo, as
discussed in Section~4. This degree of bridging is roughly consistent
with X-ray observations of clusters, where in approximately $10-20\%$
of all cases there is a significant second component roughly within
$R_{100}$, corresponding to the scale length of a $b=0.2$ FOF halo
(Vikhlinin~2007). We have found that the bridged halo fraction rises
as a function of mass, and when rescaled by the collapse mass scale
$M_*$, also appears to be universal. We also find that in the cluster
mass regime the fraction of halos with major satellites as function of
the satellite's mass fraction is cosmology dependent.

The bridged FOF halo fraction complicates the procedure for
transforming the global mass function.  Accurate mapping between the
global FOF and SO mass function must take into account SO
multiplicity within FOF halos due to the bridging (which should be
distinguished from the substructure mass function). Fortunately, if
the bridged halo fraction is universal, then this problem can be
(approximately) solved by one more iteration of the procedure
described here. A simple prescription for handling the bridging
problem, for example, may be the simultaneous use of two different
linking lengths as a way of identifying substructure in the FOF halo
identified with the longer ($b=0.2$) linking length. Then, with mock
halo mappings available for the shorter linking length, one would
construct a new mass function which should be almost free of bridging
artifacts to at least the $5\%$ level. This possibility is currently
under investigation.

In this work, systematic and statistical uncertainties were held to
$\sim5\%$, which represents the current state of the art in
determining the halo mass function. The sensitivity of halo masses to
simulation parameters such as force and mass resolution has not yet
been satisfactorily controlled below this level. While further
improvement is not ruled out, the universality of the FOF mass
function is not known to be valid at or better than this level either.

The finite bridged halo fraction points to the existence of some level
of bias when applying simple HOD schemes for the distribution of
galaxies in halos, due to the existence of (minor/major) halo
substructure.  In standard HOD methods, halos are often selected, or
assumed to be selected, by the FOF algorithm. However, this standard
method then assumes a spherically-symmetric (usually NFW) distribution
of satellite galaxies within halos, which is possibly at odds with a
significant fraction of real halos (see, e.g., Berlind \& Weinberg
2002; Tinker et al. 2005). The fraction of problematic, irregular
morphology FOF halos is mass-dependent, creating thereby a mass
dependent source of error. Furthermore, any concentration dependence
of the fraction of bridged FOF halos makes it difficult to
parameterize halo properties purely as a function of halo mass, which
is standard within HOD methods.

Despite these difficulties, the availability of sufficiently high
resolution simulations should yield a completely satisfactory HOD more
or less independent of the particular halo definition used (FOF or
SO), provided that a realistic satellite distribution is implemented.
The point is that, even with such a simulation, a simplified
description of halos such as an NFW profile for populating halos with
galaxies, would certainly fail for a not insignificant fraction of
halos, and be a cause of systematic errors.

As an alternative to mapping SO mass functions beginning with the
universal form of the FOF mass function, and utilizing the
cosmology-dependent concentration-mass relation and its scatter, one
could instead take the more computationally expensive approach of
computing SO mass functions from simulations that sample a range of
plausible cosmologies (e.g., Tinker et al.~2008). The additional
expense of such an approach can be drastically reduced by the use of
efficient statistical sampling and interpolation techniques that have
been successfully demonstrated for cosmic microwave background
temperature anisotropy and for the mass power spectrum (Heitmann et
al.~2006b, Habib et al.~2007). This work is currently in progress.

We remain agnostic as to the value of particular choices of halo
definitions and masses in cosmological applications. For X-ray
observations of relaxed clusters, the SO approach appears to be more
natural since one fits directly to a spherically averaged profile as
is observational practice. High-resolution views of the gas
distribution in clusters (e.g., Jeltema et al.~2005) are hardly
consistent with spherical symmetry, however, and the physics of the
underlying robustness of the mass-observable relations remains to be
fully established. Turning to other applications such as optical group
and cluster and subcluster member identification, there may be no
option but the use of (modified) FOF techniques.  Analagous to our
bridged FOF halos, Sunyaev-Zel'dovich observations are likely to
suffer from bridging of closely-neighboring clusters.  Mock catalogs
for ongoing and future cluster observations carried out via the
Sunyaev-Zel'dovich effect have been built using FOF definitions for
clusters (albeit with shorter linking lengths than $b=0.2$), as the
possible systematics from using spherical halo definitions are not
clear (Schulz \& White~2003).

\acknowledgements

We thank Daniel Eisenstein, Gus Evrard, Savvas Koushiappas, Beth Reid,
Paul Ricker, Alexey Vikhlinin, David Weinberg, Martin White, and Ann
Zabludoff for useful discussions. The authors acknowledge support from
IGPP, LANL, and from the DOE via the LDRD program at LANL. We are
particularly grateful for supercomputing support awarded to us under
the LANL Institutional Computing Initiative.

\end{document}